\documentclass[oldversion,rnote]{aa}
\usepackage{graphicx}
\begin{document}
\title{The narrow and moving HeII lines in nova KT Eri}
   \author{U.~Munari\inst{1},
	   E.~Mason\inst{2} and
           P.~Valisa\inst{3}
             }

   \offprints{ulisse.munari@oapd.inaf.it}

  \institute{INAF Osservatorio Astronomico di Padova, 36012 Asiago (VI), Italy
      \and INAF Osservatorio Astronomico di Trieste, 34143 Trieste, Italy
      \and ANS Collaboration, c/o Osservatorio Astronomico, 36012 Asiago (VI), Italy}

   \date{Received YYY ZZ, XXXX; accepted YYY ZZ, XXXX}

     \abstract{
We present outburst and quiescence spectra of the classical nova KT\,Eri and
discuss the appearance of a sharp HeII 4686 \AA\ emission line, whose
origin is a matter of discussion for those novae that showed a similar
component.  We suggest that the sharp HeII line, when it first
appeared toward the end of the outburst optically thick phase, comes
from the wrist of the dumbbell structure characterizing the ejecta as
modeled by Ribeiro et al.  (2013).  When the ejecta turned optically thin,
the already sharp HeII line became two times narrower and originated from the
exposed central binary.  During the optically thin phase, the HeII line
displayed a large change in radial velocity that had no counterpart in the
Balmer lines (both their narrow cores and the broad pedestals).  The large
variability in radial velocity of the HeII line continued well into
quiescence, and it remains the strongest emission line observed over the
whole optical range.
    \keywords{(stars:) novae, cataclysmic variables; (individual): KT\,Eri}
               }

   \authorrunning{U. Munari, E. Mason and P. Valisa}
   \titlerunning{HeII lines in Nova KT Eri 2009}

   \maketitle

\section{Introduction}

Nova Eri 2009, later named KT Eri, was discovered by K. Itagaki on 2009
November 25.5 UT (see CBET 2050), well past its optical maximum.  Using data
obtained by SMEI (Solar Mass Ejection Imager) on board the {\it Coriolis}
satellite, Hounsell et al.  (2011) were able to reconstruct the
pre-discovery outburst light curve, which highlights a rapid rise in
magnitude after the first detection on 2009 November 13.12 UT, a sharp
maximum reached on 2009 November 14.67 UT, after which the nova immediately
entered the rapid decline characterized by
$t_2$=6.6 days.  Preliminary reports on the early spectroscopic and
photometric evolution were provided by Ragan et al.  (2009), Rudy et al. 
(2009), Bode et al.  (2010), Imamura and Tanabe (2012), and Hung, Chen, \&
Walter (2012).  Radio observations were obtained by O`Brien et al.  (2010)
and X-ray observations by Bode et al.  (2010), Beardmore et al.  (2010), and
Ness et al.  (2010).  Raj, Banerjee \& Ashok (2013) discussed early infrared
photometric and spectroscopic evolution, while the line profiles and their
temporal evolution were modeled in detail by Ribeiro et al.  (2013). 
Jurdana-{\v S}epi{\'c} et al.  (2012) searched the Harvard plate archive and
measured the progenitor of the nova on 1012 plates dating from 1888 to 1962. 
No previous outburst was found.  The photometric evolution of KT Eri after
it returned to quiescence and its persistent P=752 day periodicity
have been discussed by Munari and Dallaporta (2014).

  \begin{figure*}
     \centering
     \includegraphics[height=18cm,angle=270]{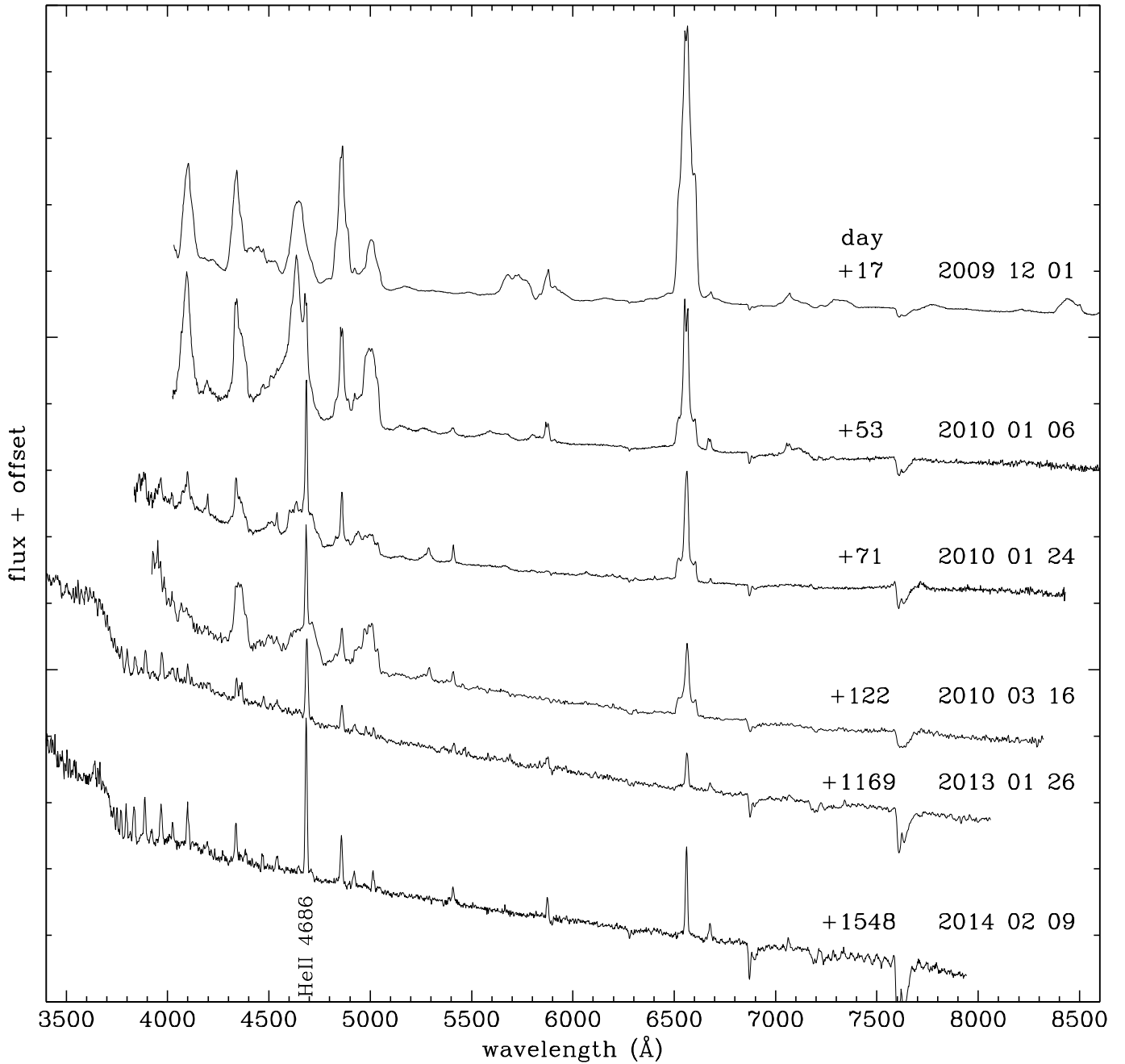}
     \caption{Sample spectra from our monitoring to highlight the
     spectroscopic evolution of KT Eri during the 2009/10 nova outburst 
     and the subsequent return to quiescence.}
     \label{fig1}
  \end{figure*}

Here we present KT\,Eri spectra taken from outburst maximum to subsequent
quiescence and focus on the appearance and evolution of a narrow HeII 4686
\AA\ emission line.  Sharp emission lines superimposed to much broader
emission components, have been observed in a few other recent novae:
YY\,Dor, nova LMC 2009, U\,Sco, DE\,Cir, and V2672\,Oph (see, e.g., the Stony
Brooks SMART Atlas\footnote{\sf
www.astro.sunysb.edu/fwalter/SMARTS/NovaAtlas/atlas.html}).  Complex line
profiles have always been modeled with axisymmetric ejecta geometries
consisting of bipolar lobes, polar caps, and equatorial rings (starting with
Payne-Gaposchkin in 1957).  Using a similar approach, the sharp and strong
narrow emission in V2672\,Oph could be successfully modeled as coming from
an equatorial ring whereas the broader pedestal originates from polar cups
(Munari et al.  2011).  However, because of their sharpness, profile, and width
it has been also suggested that the narrow components in the above systems
might arise from the accretion disk of the underlying binary (Walter \&
Battisti 2011, but see also Mason \& Walter 2013), once the ejecta becomes
sufficiently transparent.  In the case of U\,Sco, the observation of radial
velocity motion of the narrow HeII emission has been interpreted as restored
accretion shortly after the nova 2010 outburst (Mason et al.  2012). 
Whether a narrow emission component, and in particular the appearance of the
sharp HeII$\lambda$4686 line, always originates in the central binary and
recovered accretion from the secondary star has to be established.

We believe KT Eri offers an interesting bridge between these two alternative
views.  We will show how when first seen in emission in the spectra of KT
Eri, during the optically thick phase, the sharp HeII 4686~\AA\ line
was coming from the inner and slower regions of the ejecta, and how, at
later times when the ejecta turned optically thin, the HeII line
became two times sharper and variable in radial velocity, indicating it was
coming directly from the central binary.  Thus, the presence and the origin
of sharp HeII emission lines seems to depend on the geometry of the ejecta,
their viewing angle, and on the evolutionary phase of the nova.

\section{Observations}

Absolute spectroscopy of KT Eri was obtained during the outburst as part of
the long-term ANS Collaboration monitoring of novae in eruption (see Munari
et al.  2012).  We used the Varese 0.61m telescope equipped with the mark.II
Multi Mode Spectrograph (Munari and Valisa 2014), that allows rapid
switching between low-dispersion, medium-dispersion and Echelle high-resolution
modes.  A 2 arcsec wide slit, aligned along the parallactic angle was used
in all observations.  Echelle spectra were calibrated with a thorium lamp
exposed before and after the science spectra, and similarly with a FeHeAr
lamp for the medium- and low-dispersion spectra.  Flux calibration was
achieved via observation of the nearby spectrophotometric standard HR 1784
observed with the identical instrumental setup both immediately before and
soon after the nova.  All data reduction was carried out in IRAF, following
standard extraction procedures involving correction for bias, dark, and flat
and sky background subtraction.

Low-resolution absolute spectroscopy of KT Eri, after it returned to
quiescence brightness long after the end of the outburst, was obtained with
the Asiago 1.22m telescope and B\&C single dispersion spectrograph.  Also in
this case we adopted a 2 arcsec wide slit, aligned along the parallactic
angle, the same spectrophotometric standard star and the same
extraction/calibration procedures as for the observations obtained with the
0.61m telescope during the outburst phase.

Table~1 provides a logbook of the spectroscopic observations. The
visibility and integrated flux of the HeII 4686 \AA\ emission line is given
in Table~2.  The values listed in Table~2 are averaged over all the spectra
collected on the given observing night.  The heliocentric radial velocity of
the HeII 4686 \AA\ emission line, as measured on each individual collected
spectrum, is listed in Table~3.  The radial velocities are from Gaussian
fits to the observed profiles.  The observed HeII profile rapidly converged
to a Gaussian-like profile shortly after the line appeared in emission.  The
profile at the earliest epochs was admittedly more structured than a simple
Gaussian, but it was basically symmetric and therefore the fit with a
Gaussian does not significantly affect the derived velocity, which is
very similar to the velocity derived for the line photocenter.  The
typical measurement error for these Gaussian fits is 12 km~s$^{-1}$.  Any
error in the wavelength calibration of the spectra is removed from the HeII
radial velocities by subtracting from them the average velocity measured, on
the sky background, for the two HgI city-light emission lines at 4358 and
5461 \AA.  These corrections are on the order of 10 km~s$^{-1}$
(corresponding to 7\% of the velocity span of one pixel).

  \begin{table}[!Ht]
     \centering
     \caption{Logbook of our spectroscopic observations of KT Eri.
     Resolving power is listed for the Echelle high-resolution observations,
     dispersion for the medium- and low-resolution spectra.  $\Delta
     t$ is the time past the optical maximum as fixed by SMEI observations
     (2009 November 14.67 UT or HJD=2455150.17; Hounsell et al.  2011).}
     \includegraphics[width=8.8cm]{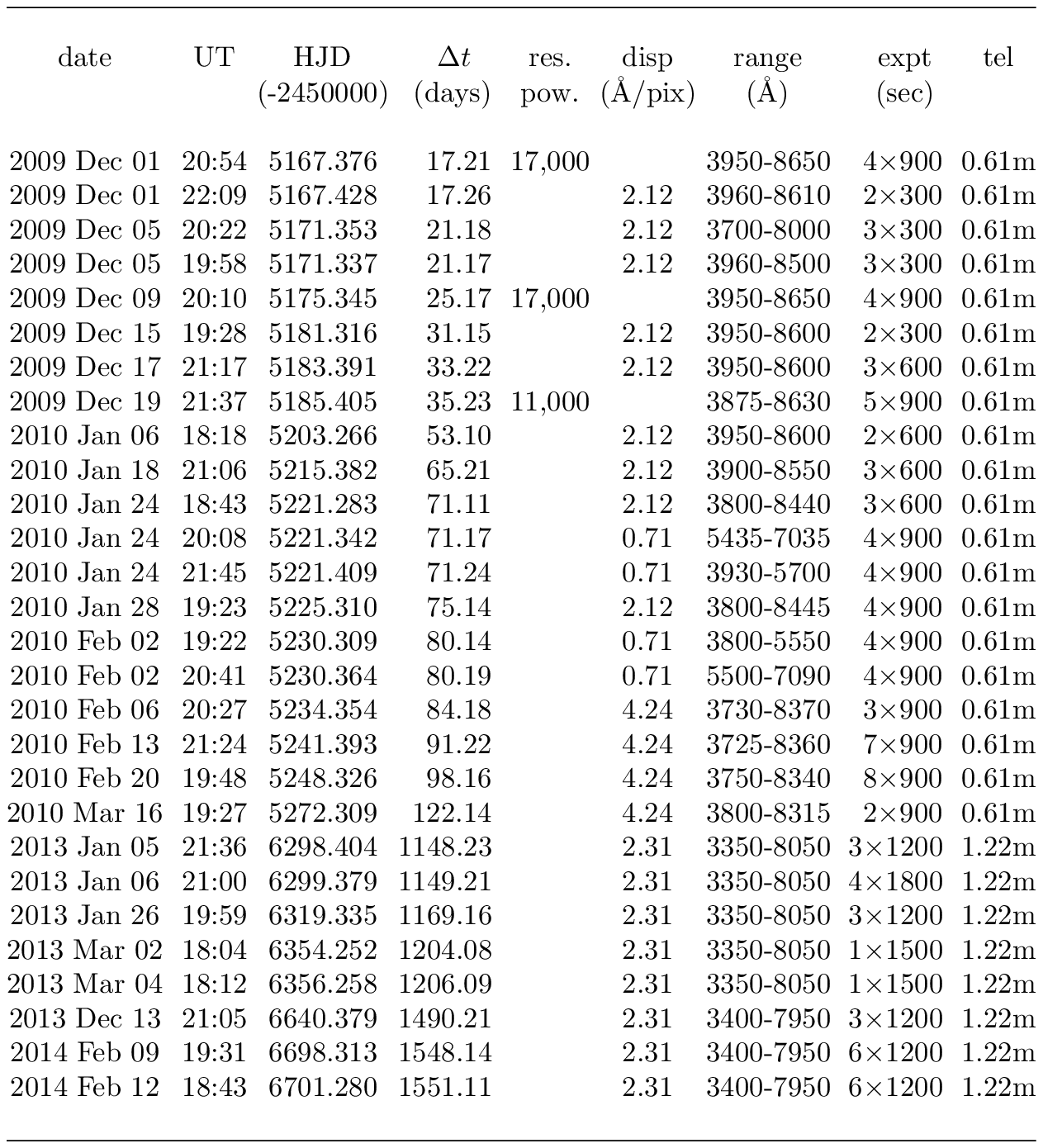}
     \label{tab1}
  \end{table}

  \begin{table}[!Ht]
     \centering
     \caption{Visibility and integrated absolute flux
     of HeII 4686 \AA\ emission line in our spectra of KT Eri.}
     \includegraphics[width=8.8cm]{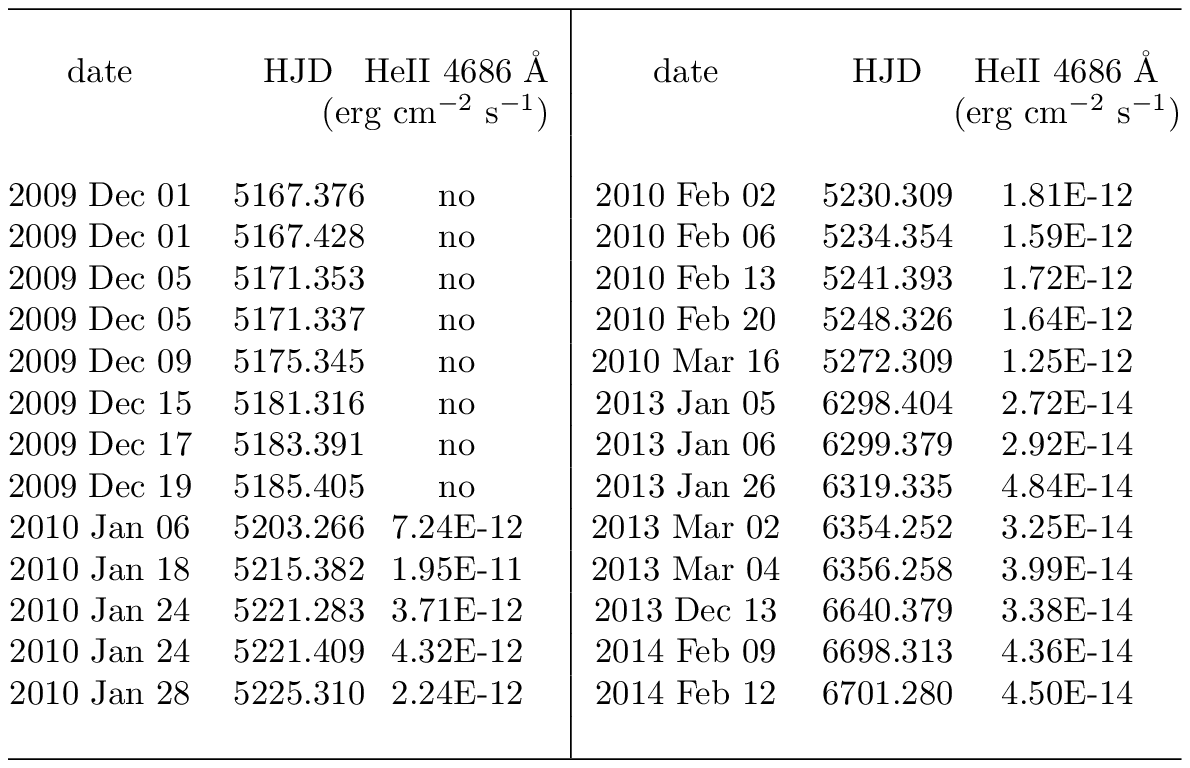}
     \label{tab2}
  \end{table}

  \begin{figure}[!Ht]
     \centering
     \includegraphics[width=8.8cm]{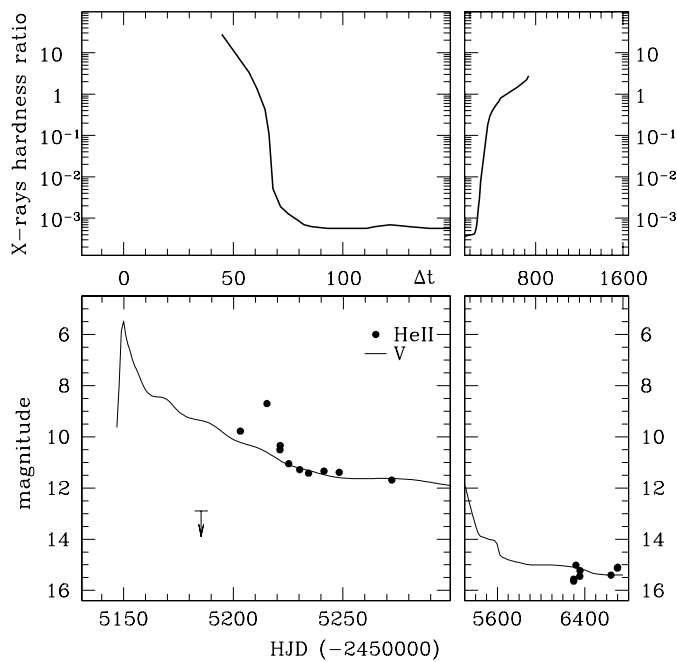}
     \caption{Evolution of the integrated flux of HeII 4686 \AA\ emission
     line of KT Eri compared to its $V$-band light curve (from Hounsell et
     al.  2010, AAVSO database, Munari \& Dallaporta 2014, and unpublished
     recent data).  To facilitate the comparison, the integrated line flux is
     transformed in magnitudes and offset by the quantity given in Eq.  (1). 
     The arrow marks the upper limit to HeII integrated flux on day +35
     spectra. Top panel: the hardness ratio (defined
     as the ratio between the count rates in the 1-10 keV and 0.3-1 keV
     bands) of the Swift X-ray observations (adapted from public data
     available on the Swift web site).}
     \label{fig2}
  \end{figure}

  \begin{figure}[!Ht]   
     \centering
     \includegraphics[angle=270,width=8.8cm]{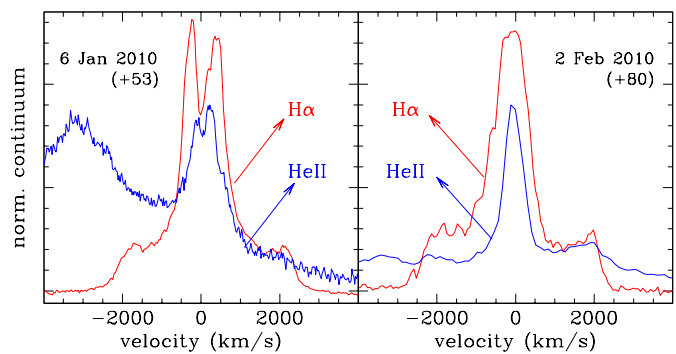}
     \caption{Comparison between the line profiles of H$\alpha$ and HeII
     4686 \AA\ for day +53 (ejecta still optically thick) and day +80
     (ejecta now optically thin).}
     \label{fig3}
  \end{figure}

\section{Spectral evolution and HeII lines}

The spectral evolution of KT Eri during the 2009 outburst and the subsequent
return to quiescence is presented in Figure~1.  The average pre-outburst
mean $B$ magnitude of KT Eri was around 15.4 mag (from Jurdana-{\v
S}epi{\'c} et al.  2012, and the recalibration of their historical Harvard
photographic data as performed by Munari and Dallaporta 2014).  The average
$B$-band magnitude of KT Eri during 2013 and 2014 is 15.3, which confirms
that the object was back to quiescence level when we observed it in 2013 and
2014.

The spectrum for 2009 December 01 (day +17) is representative of those
obtained immediately following the discovery of the nova (that happened +11
days past optical maximum).  It is characterized by broad emission lines. 
The average FWHM of hydrogen Balmer and OI emission lines is 3200
km~s$^{-1}$, while HeI lines are sharper with a FWHM of 950 km~s$^{-1}$. 
The profile of the Balmer lines already shows hints of the two components - a
broad pedestal and a superimposed sharper core - that will stand out clearly
at later epochs.  No HeII 4686 \AA\ emission line was observed in our early
spectra, including the Echelle for 2009 December 19, +35 day past maximum. 
Our next spectrum, for 2010 January 6, corresponding to day +53, shows a
weak HeII in emission superimposed on the red wing of the strong NIII 4640
\AA\ line.  The exact day of the first appearance of HeII is day +48, as
revealed by inspection of the compilation of spectra of KT Eri obtained with
the Liverpool telescope + FRODOS spectrograph and summarized by Ribeiro
(2011; one of these spectra is used to plot the left panel of Figure~3 in
place of our spectrum for the same date that was obtained at a lower
resolution).  The sample of spectra in Figure~1 shows how the HeII line
rapidly grows in equivalent width with the progress of the decline from
maximum, and how HeII 4686 \AA\ remains the strongest spectral feature
during the subsequent phases of the outburst and the following return to
quiescence.  

  \begin{figure*}[!Ht]
     \centering
     \includegraphics[angle=270,width=18.5cm]{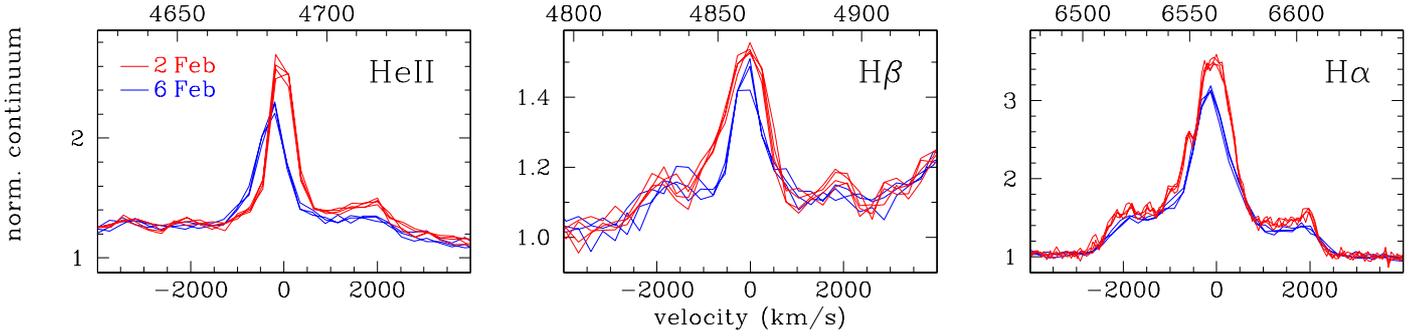}
     \caption{Overplot of HeII 4686 \AA, H$\beta$, and H$\alpha$ lines from
     individual spectra obtained on 2010 February 2 and 6 (rebinned to the
     same, coarser wavelength scale of the later date).  It is quite obvious
     how the large radial velocity change displayed by HeII between the two
     dates does not have a counterpart in the Balmer lines, neither the
     broad pedestal nor the narrower core.}
     \label{diffs}
  \end{figure*}

The flux evolution of the HeII 4686 \AA\ emission line is compared to the
$V$-band light curve of KT Eri in Figure~2. To transform the flux of HeII
4686 \AA\ into a magnitude scale for an easier comparison with the nova
evolution in the $V$-band, we computed its {\it magnitude} as
\begin{equation}
{\rm mag}{\rm (HeII)} = -2.5 \times \log \left( \frac{flux}{5.9 \times 10^{-8}}
\right),
\end{equation}  
where the arbitrary constant $5.9 \times 10^{-8}$ erg cm$^{-2}$ s$^{-1}$ is
chosen so to cancel the shift in Figure~2 between the $V$-band and the HeII
light curves.  Figures~2 aims to highlight two basic facts: ($a$) the HeII
line appeared when the nova had declined by about 3.5/4.0 mag below maximum
brightness, i.e.,  a characteristic time in the evolution of typical novae
when the ejecta begin turning optically thin allowing direct vision of the
central star (e.g.,  McLaughlin 1960, Munari 2012).  The fact that the ejecta
were becoming optically thin at that time is confirmed by the simultaneous
huge increase in the soft component of the X-ray emission.  The X-ray
hardness ratio from Swift observations is plotted in the top panel of
Figure~2 that shows how the nova entered the so-called super-soft-source
phase (SSS, Krautter 2008) around day +50, simultaneously with the
appearance of HeII in the optical spectra; ($b$) after an initial surge in
the intensity of the HeII line, by day +70 its integrated flux declined in
pace with the decline of the nova in $V$-band.  The proportionality of HeII
flux and $V$-band brightness continued well into the quiescence phase.

  \begin{table}[!Ht]
     \centering
     \caption{Heliocentric radial velocity (RV$_\odot$) of HeII 4686 \AA\
     emission line measured on our individual spectra of KT Eri. HJD =
     heliocentric JD - 2450000.}
     \includegraphics[width=8.0cm]{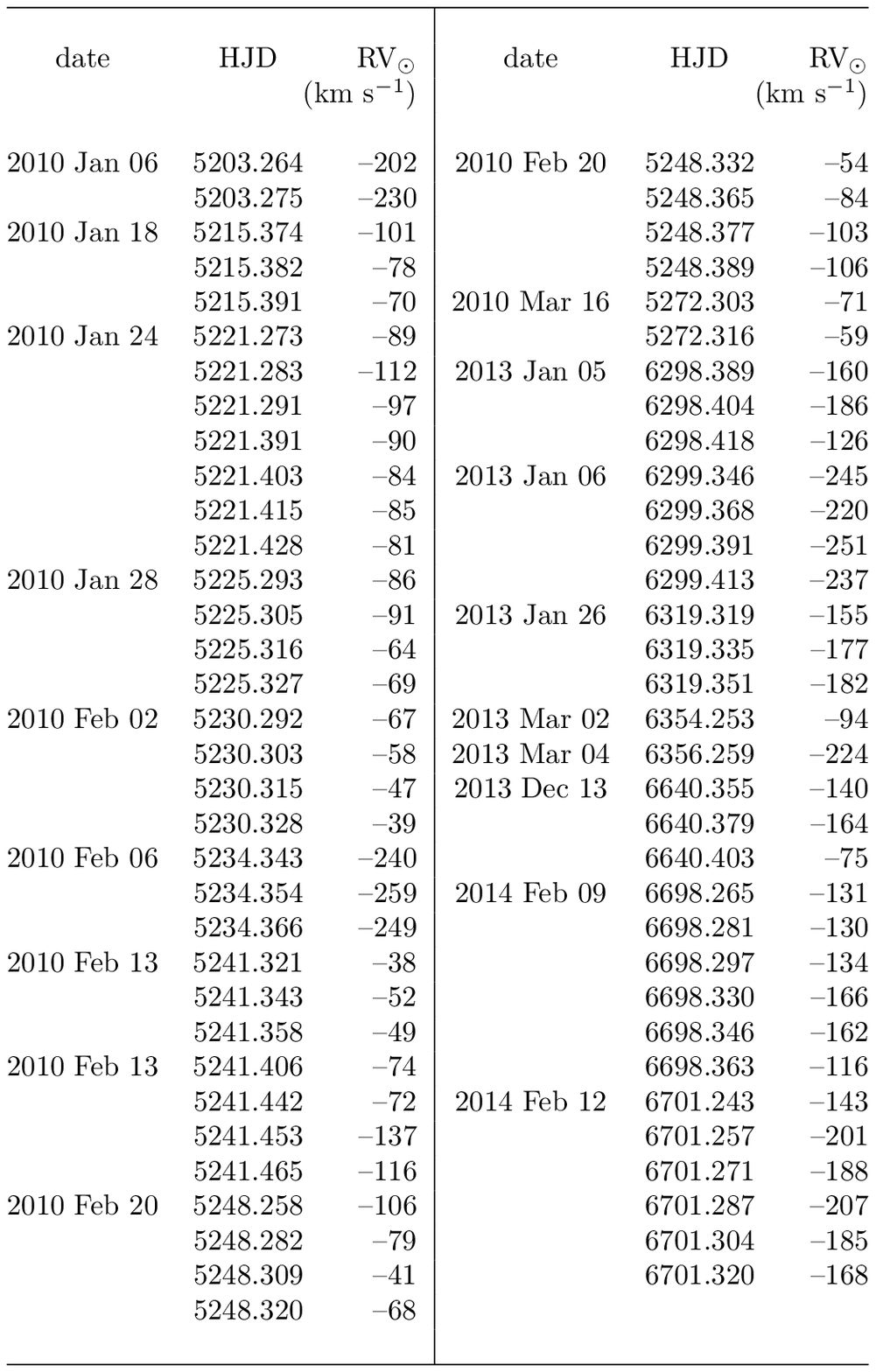}
     \label{tab2}
  \end{table}

The transition around day +70 in the flux evolution of HeII 4686 \AA\ also 
marked a conspicuous change in its profile, which is well illustrated in
Figure~3.  Day +70 also marks the time when the X-ray emission was reaching
its maximum and initiating the SSS-plateau phase characterized by the lowest
value of the hardness ratio (see top panel of Figure~2).

Before day +70 (left panel in Figure~3), HeII 4686 \AA\ was similar to the
narrow component of the Balmer lines. Both were double peaked and of
similar width: 900 km~s$^{-1}$ for HeII 4686 \AA and 1150 km~s$^{-1}$ for 
the Balmer lines. The width of the broad pedestal of the Balmer lines was 
4500 km~s$^{-1}$.

After day +70 (right panel in Figure~3), the FWHM of HeII 4686 \AA\ suddenly
dropped by a factor of two, down to 460 km~s$^{-1}$, while that of the Balmer
lines remained around 1000 km~s$^{-1}$ for the narrow component and 4600
km~s$^{-1}$ for the pedestal.

Ribeiro et al.  (2013) modeled the H$\alpha$ profile during the early
optically thick phase of KT Eri and found a good fit with dumbbell shaped
expanding ejecta and no need for an equatorial ring.  The dumbbell structure
was characterized by a 1/$r$ radial density profile, $V_{\rm exp}$ = 2800
$\pm$ 200 km~s$^{-1}$, a major to minor axis ratio of 4:1, and an
inclination angle of 58$_{-7}^{+6}$ deg.  The density profile $\rho \propto
r^{-1}$ allowed Ribeiro et al.  to fit the broad, square-like pedestal
coming from the outer parts of the bipolar lobes and the narrow,
double-peaked central component of the H$\alpha$ profile coming from the
slower and denser regions closer to the wrist, simultaneously.

When HeII 4686 \AA\ was first weakly detected around day +48, and for the
following period up to day +70, it came from the denser region of the ejecta
closer to the wrist of the bipolar structure.  This is the same region from
where at earlier times the HeI lines came from.  In fact, their FWHM (950
km~s$^{-1}$) was very similar to that of HeII 4686 \AA\ and the narrow
component of the Balmer lines.  The appearance of HeII 4686 \AA\ (produced
during the recombination of HeIII to HeII) was obviously related to the
increase in temperature of the pseudo-photosphere contracting through the
inner regions of the ejecta closer to the central star.  As this contraction
proceeded and the optical thickness of the ejecta continued to decline as a
consequence of the ongoing expansion, a larger fraction of the inner ejecta
were reached by hard ionizing photons and the intensity of HeII 4686 \AA\
surged until a maximum was attained around day +65.  The ionization of HeII
into HeIII never reached the outer lobes of the ejecta, because the HeII
lines never developed the broad pedestal displayed by the Balmer
lines\footnote{The apparent emission bump on the red side of the narrow HeII
line in the left panel of Figure~4 should be identified with other
transitions within the blend, as, otherwise, we should observe a symmetric
component on the blue side of HeII}.  Following this maximum, the intensity
of HeII 4686 \AA\ began to decline in parallel to the other emission lines
and to the underlying continuum.

Day +70 also marks the time when the continuum emission from the ejecta, now
completely transparent, fell below that coming directly from the central
star.  As shown in Figure~2, by day +70 ($a$) the X-ray emission entered the
SSS plateau where it remained stable until day $\sim$250 which marks the end
of the nuclear burning on the white dwarf, and ($b$) in parallel, the optical
brightness of the KT~Eri stopped declining because direct emission from the
white dwarf and from the irradiated companion replaced that of fading
ejecta; it rapidly dropped to the quiescence value only when the X-ray SSS
phase ended.  The broader HeII 4686 \AA\ profile (FWHM 900 km~s$^{-1}$)
coming from the fading ejecta after day +70 is overwhelmed by the narrower
profile originating directly from the central binary (FWHM 460 km~s$^{-1}$). 
This will not happen for the Balmer lines until much later into the evolution,
because (as shown by the quiescence spectrum from 2013 January 26, day
+1169) the Balmer lines produced by the central star are quite weak and to
emerge they need the emission from the ejecta to essentially vanish.  

The sequences of spectra in Figure~4 show how - past day +70 - the HeII 4686
\AA\ emission originated directly from the central binary while the Balmer lines
continue to come from the ejecta for a long time.  Here the profiles of
H$\alpha$, H$\beta$, and HeII 4686 \AA\ from many different spectra obtained
on 2010 February 2 and 6 (days +80 and +84) are compared.  The
profile and radial velocity of the Balmer lines, both their broad pedestal and
their narrow component, do not change from one night to the other.  The
$\sim$200 km~s$^{-1}$ shift in radial velocity of HeII (from -53 to -249
km~s$^{-1}$, see Table~3) is instead outstanding.  Such continuous
and large change in radial velocity of HeII 4686 \AA\ cannot be easily
understood in terms of ballistic expansion of nova ejecta, while they can be
naturally accounted for by the continuously changing viewing geometry of the
central binary and the instabilities inherent to mass transfer.

We have searched the epoch radial velocities in Table~3 looking for
periodicities that could betray the orbital period of the nova.  We have
used the Fourier code implemented by Deeming (1975) for unequally spaced
data.  Extensive tests on the whole set of Table~3 data as well as on random
selected subsamples failed to reveal any clear and strong periodicity.  The
situation is similar to that encountered in photometry (see Munari and
Dallaporta 2014), where no other clear periodicity stands out in addition to
the 752-day eclipse-like events that regularly marked the pre- and
post-outburst optical photometry of KT Eri.  The brightness of KT Eri varies
similarly in all optical bands and by a large amplitude (on the order of one
magnitude in $B$) with a timescale that is continously changing, in an
apparently chaotic manner.  The same seems to occur with the radial
velocity of the HeII emission line, which is seen to change by large margins,
but in an apparently chaotic pattern.  This could be the result of the
beating of several different true periodicities simultaneously present, but
to disentangle them it will be necessary to accumulate many additional
observations, possibly at a higher dispersion than the spectra used for the
present study.  Additional data, regularly spaced over many consecutive
observing seasons, will be necessary to investigate the presence of the 
752-day period among the spectral data.

\begin{acknowledgements}
We would like to thank A. Milani and V. Luppi of ANS Collaboration for their
assistance with the observations obtained with the Varese 0.61m telescope and
Valerio A.R.M.  Ribeiro for having promptly offered his higher quality
spectrum for 6 Jan 2010.  EM thanks Steven N. Shore for valuable discussions.

\end{acknowledgements}

\end{document}